\UseRawInputEncoding
\documentclass[prd,amsmath,amssymb,showpacs,superscriptaddress,nofootinbib]{revtex4-1}
\usepackage{graphicx}
\usepackage{epsfig,graphics,subfigure,psfrag,amsmath,amssymb}
\usepackage{lineno}
\usepackage{dcolumn}
\usepackage{bm}
\usepackage{overpic}
\usepackage{xspace}
\usepackage{xcolor}
\usepackage{rotating}
\usepackage{color}
\usepackage{multirow}
\usepackage{colortbl}
\usepackage{epstopdf}
\usepackage{float}
\usepackage{makecell}
\usepackage[colorlinks,linkcolor=blue,anchorcolor=blue,citecolor=blue]{hyperref}
\usepackage{setspace}
\newcommand{\ra}{\rightarrow}
\newcommand{\no}{\nonumber}

\newcommand{\jp}{J/\psi}

\newcommand{\gev}{\mathrm{~GeV}}

\begin{document}
\title{\boldmath Study of the  $\jp\to \Lambda\bar{\Sigma}^0\eta$ reaction}
\author{
  \begin{center}
L. R. Dai$^{1,3}$,
Wen-Tao Lyu$^{2,3}$,
E. Oset$^{3,4}$
\\
\vspace{0.2cm} {\it
$^{1}$ School of science, Huzhou University, Huzhou, 313000, Zhejiang, China\\
$^{2}$ School of Physics, Zhengzhou University, Zhengzhou 450001, China\\
$^{3}$ Departamento de F¨ªsica Te¨®rica and IFIC, Centro Mixto Universidad de Valencia-CSIC Institutos de Investigaci¨®n de Paterna, 46071 Valencia, Spain\\
$^{4}$ Department of Physics, Guangxi Normal University, Guilin 541004, China
}
\end{center}
}

\date{\today}

\begin{abstract}
We study the isospin violating $\jp \to \Lambda\bar \Sigma^0 \eta $ reaction, recently measured by the BESIII collaboration, by looking at the dominant terms with $\bar \Sigma^0$  and a pair of pseudoscalar-baryon particles that form together an SU(3) singlet and can thus couple to the $\jp$. Next we allow these pairs to undergo final state interaction to produce the final $\eta \Lambda$.  We find that the relevant original channels are $\bar K N$ and $K \Xi$,  and the non cancelation of terms involving charged and neutral particles, because of their different masses, is responsible for the reaction. With that mechanism we find a good agreement with the three experimental mass distributions.

\end{abstract}

\maketitle

\section{INTRODUCTION}

The BESIII collaboration has reported recently results on the $\jp \to \Lambda \bar \Sigma^0 \eta +\mathrm{c.c}$ reaction \cite{besexpe}, where
a clear signal for the $\Lambda(1670)$ excitation is seen in the $\eta \Lambda$  mass distribution, which accounts for practically all the
strength of this magnitude. The reaction violates isospin, which offers a nice opportunity to investigate the origin of isospin violation, a relevant issue in the strong interaction, which conserves this magnitude.

 The $\Lambda(1670)$  lies very close to the $\eta \Lambda$ threshold which makes it a candidate to be a molecular state, although the closeness to a threshold, while favoring molecular components \cite{bonnmole,guocarlos,hyodo} is not a sufficient condition to have a state of molecular  nature \cite{daisong1,daisong2,haipengsong}.

 But, since long, the  $\Lambda(1670)$ has been identified as a dynamically generated, together with the two $\Lambda(1405)$ states, coming from the interaction of the coupled channels $\bar{K} N$, $\pi \Sigma$, $\pi \Lambda$,
$\eta \Sigma$, $\eta \Lambda$, $K \Xi$ \cite{angels,bennhold}. The $\Lambda(1670)$ was found in \cite{bennhold} to correspond to a state coupling to $K \Xi$ with
large strength, next to  $\eta \Lambda$, then to  $\bar{K} N$, and coupling very weakly to $\pi \Sigma$.

 The nature of a resonance as a dynamically generated state from  coupled channels interaction opens a door for a larger  isospin violation than
 expected for elementary particles in reactions involving these  states. This is indeed the case in the small width for the $D^*_{s0}(2317)$
resonance, due to the isospin forbidden decay into $D_s \pi^0$. Indeed, the $D^*_{s0}(2317)$ width, assuming the $D^*_{s0}(2317)$ to be an ordinary state, is much smaller than considering it a molecular state, mostly due to the $KD$ interaction \cite{39,26,41,42,43,arxiv,dani,iketole}. To understand
the reason one must go back to early works \cite{44}, showing that there is an important source of isospin violation from
loops containing  kaons, due to the different masses of the charged and neutral kaons.
The kaonic loops have been identified as the source of  isospin violation  in many reactions \cite{achasov,achakise,achashes,kubispela,hanhartpela, rocasolo,acetizou,sakailiang,jiaxinjiang,jiatingli}.
In the case of  the $\Lambda(1670)$ state, one has loops from $K^- p$, $\bar{K}^0 n$ but also loops from $K^+ \Xi^-$, $K^0 \Xi^0$ which also have different masses, and it would be interesting to see how these channels influence the isospin violation in the $\jp\to \Lambda\bar{\Sigma}^0\eta$  decay. This is the purpose of the present work, where we will show that the influence of the $K \Xi$ channels  is much more important  than that  of the $\bar{K} N$ channels.

The  $\jp\to \Lambda\bar{\Sigma}^0\eta$ reaction can be considered  a complement to two other reactions,
$\Lambda^+_c \to \pi^+ K^- p$ \cite{ bellelamb} and $ K^- p \to \eta \Lambda$ \cite{starostin},  which were studied in \cite{manyu}, from where  the   $\eta \Lambda$ scattering length and effective range were extracted. We shall use here  the same input  used in the study  of the
$\Lambda^+_c \to \pi^+ K^- p$ reaction, and the agreement obtained  with the data here corroborates the findings and conclusions obtained in \cite{manyu}, among others, the precise values of $a,r_0$ extracted for $\eta \Lambda$ scattering  close to threshold in \cite{manyu}.

\section{formalism}
   The idea is that we start with the strong decay of $\jp$ into $\bar{\Sigma}^0 B P$, where $B$ is a baryon of the    octet of $1/2^+$ state and $P$ a pseudoscalar of the $0^-$ octet. The $\bar{\Sigma}^0 B P$ system originally satisfies isospin symmetry, which implies that $BP$  has isospin $I=1$, then the $BP$  interaction leads us
   finally  to $\eta \Lambda$ violating isospin, which occurs through loops not cancelling  because of the different  masses of $BP$ belonging  to the same  isospin multiplet.

     To see which $BP$  states are produced in the $\jp\to \Lambda\bar{\Sigma}^0\eta$ reaction we consider that
  $\jp$ is an SU(3) singlet  in the $u,d,s$ sector. Then writing matrices  for $q\bar{q}$ states of the octet in
terms of mesons and $qqq$ for baryons of the octet we have

\begin{eqnarray}\label{eq:Pmatrix}
   P=
    \left(
    \begin{array}{ccc}
    \frac{1}{\sqrt{2}}\pi^0 + \frac{1}{\sqrt{3}} \eta & \pi^+ & K^+ \\[2mm]
    \pi^- & -\frac{1}{\sqrt{2}} \pi^0 + \frac{1}{\sqrt{3}} \eta & K^0 \\[2mm]
    K^- & \bar{K}^0 & ~-\frac{1}{\sqrt{3}} \eta\\
    \end{array}
    \right)\,, \quad
   B=
    \left(
    \begin{array}{ccc}
    \frac{1}{\sqrt{2}}\Sigma^0 + \frac{1}{\sqrt{6}} \Lambda & \Sigma^+ & p\\[2mm]
    \Sigma^- & -\frac{1}{\sqrt{2}} \Sigma^0 + \frac{1}{\sqrt{6}} \Lambda & n \\[2mm]
    \Xi^- & \Xi^0 & ~-\frac{2}{\sqrt{6}} \Lambda\\
    \end{array}
    \right) \,, \no
\end{eqnarray}
\begin{eqnarray}
   \bar{B}=
    \left(
    \begin{array}{ccc}
    \frac{1}{\sqrt{2}}\bar{\Sigma}^0 + \frac{1}{\sqrt{6}} \bar{\Lambda} & \bar{\Sigma}^+ & \bar{\Xi}^+ \\[2mm]
    \bar{\Sigma}^- & -\frac{1}{\sqrt{2}} \bar{\Sigma}^0 + \frac{1}{\sqrt{6}} \bar{\Lambda }&  \bar{\Xi}^0\\[2mm]
    \bar{p} & \bar{n} & ~-\frac{2}{\sqrt{6}} \bar{\Lambda}\\
    \end{array}
    \right)\,.
\end{eqnarray}

There are many ways to construct SU(3) singlet structures with these matrices, calculating the trace of products, as
$\langle \bar{B} B P \rangle$, $\langle \bar{B} P B \rangle$,  $\langle \bar{B} B\rangle \langle P \rangle$,
$\langle \bar{B} P\rangle \langle B \rangle$, $\langle B P\rangle \langle \bar{B} \rangle$,
$\langle  \bar{B} \rangle \langle  B \rangle \langle  P \rangle$. However, using elements of heavy quark symmetry, one finds that the largest
contribution comes from the structures containing a smaller number of traces \cite{manohar,abreudai}, and we will stick to those terms only.
Thus, we consider the $\langle \bar{B} B P \rangle$ and $\langle \bar{B} P B \rangle$ structures in our study.  We find
\begin{eqnarray}
\langle \bar{B} B P \rangle = \frac{1}{\sqrt{2}}\bar{\Sigma}^0  \left\{ 2 \frac{1}{\sqrt{2}}\bar{\Sigma}^0 \frac{1}{\sqrt{3}} \eta +
2 \frac{1}{\sqrt{6}} \Lambda \frac{1}{\sqrt{2}} \pi^0 + \Sigma^+ \pi^- + p K^{-} -\Sigma^- \pi^+ - n \bar{K}^0
 \right\}\,.
\end{eqnarray}
However, since it was found that in \cite{bennhold}  that the $\pi\Sigma$ states couple very weakly to the $\Lambda(1670)$, we neglect these
channels and we find
\begin{eqnarray} \label{eq:3}
\langle \bar{B} B P \rangle = \frac{1}{\sqrt{2}}\bar{\Sigma}^0  \left\{ 2 \frac{1}{\sqrt{2}}\bar{\Sigma}^0 \frac{1}{\sqrt{3}} \eta +
2 \frac{1}{\sqrt{6}} \Lambda \frac{1}{\sqrt{2}} \pi^0 + p K^{-} - n \bar{K}^0
 \right\}\,.
\end{eqnarray}

With the isospin phase convention of our approach $(K^+,K^0),(\bar{K}^0,-K^-)$ the $p K^{-} - n \bar{K}^0$ combination  in Eq.~\eqref{eq:3}
has $I=1$, as it should, such that combined with the $\bar{\Sigma}^0$ can give  the $I=0$ of the $\jp$. The $\Sigma^0 \eta$  and  $\Lambda \pi^0$
have both $I=1$ and can not make a transition  to $\eta\Lambda$ with a strong interaction. However, the   $K^{-} p$ and $n \bar{K}^0$ can lead
to $I=0$ since the loops  of $K^{-} p$ and $n \bar{K}^0$ involved in the final state interaction will not cancel. For  practical purposes we can just take as the object produced in the first step of the $\jp$ decay
\begin{eqnarray} \label{eq:4}
\langle \bar{B} B P \rangle \ra  \frac{1}{\sqrt{2}}\bar{\Sigma}^0 (p K^{-} - n \bar{K}^0)\,.
\end{eqnarray}

 Now we look at the $\langle \bar{B} P B  \rangle $ structure. We obtain
\begin{eqnarray}
\langle \bar{B} P B  \rangle=\frac{1}{\sqrt{2}}\bar{\Sigma}^0  \left\{\frac{2}{\sqrt{12}}  \pi^0  \Lambda  + \frac{2}{\sqrt{6}} \eta \Sigma^0
+ \pi^+ \Sigma^- + K^+ \Xi^- - \pi^- \Sigma^+ -K^0 \Xi^0 \right\}
\end{eqnarray}
and neglecting the  $\pi \Sigma$ channels as before,  and the $\pi^0  \Lambda$ and   $ \eta \Sigma^0$ which cannot make transition to an $I=0$ state, we obtain
\begin{eqnarray} \label{eq:6}
\langle \bar{B} P B  \rangle \ra  \frac{1}{\sqrt{2}}\bar{\Sigma}^0 ( K^+ \Xi^-  - K^0 \Xi^0)
\end{eqnarray}

Once again, considering the isospin doublet $(\Xi^0,-\Xi^-)$ in analogy to $(\bar{K}^0,-K^-)$, we see that the   $K \Xi$ combination
 of  Eq.~\eqref{eq:6} has $I=1$, as it should. Yet, in the final state interaction to produce  $\eta\Lambda $, the  $ K^+ \Xi^- $
and $K^0 \Xi^0$ loops will not cancel and we shall have an isospin  violating mechanism, which can produce  $\eta\Lambda$ at the end, with
the shape of the $\Lambda(1670)$.

\begin{figure}[h!]
\centering
\includegraphics[scale=.7]{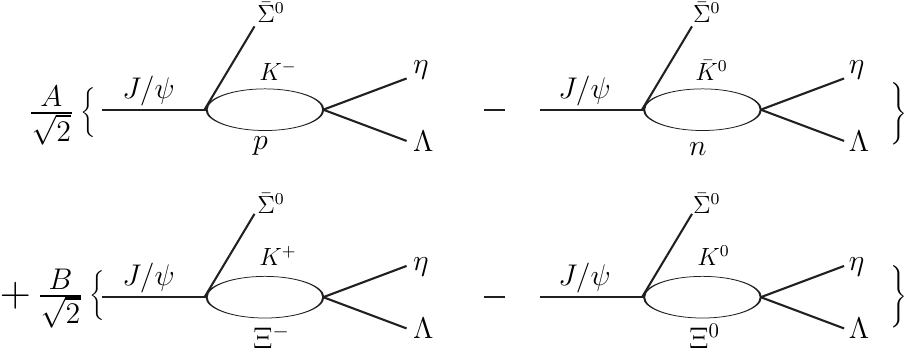}
\caption{Diagrams involving final state interaction of $PB$ pairs leading to  the   $\jp\to \Lambda\bar{\Sigma}^0\eta$   decay.}
\label{fig:1}
\end{figure}

 The two  structures   $\langle \bar{B} P B  \rangle$  and $\langle \bar{B} B P \rangle$ are different, as we see in
  Eqs.  \eqref{eq:4},\eqref{eq:6}, and we give them a different weight, $A, B$ respectively.  Thus we have for practical purposes
 \begin{eqnarray}
 \jp \ra \frac{1}{\sqrt{2}}\bar{\Sigma}^0 \left\{A (p K^{-} - n \bar{K}^0) + B ( K^+ \Xi^-  - K^0 \Xi^0) \right\}
\end{eqnarray}
 Then the transition amplitude  $\jp\to \Lambda\bar{\Sigma}^0\eta$ is given by  the diagrams of Fig.~\ref{fig:1},

The interaction of coupled channels leading to the formation of the $\Lambda(1670)$ is in $S$-wave, hence we  keep the pair  $BP$ in
$S$-wave, but $\bar{\Sigma}^0 \eta \Lambda$ has the combination $J^P, \frac{1}{2}^- 0^- \frac{1}{2}^+$, which has positive parity,  hence
we need a $P$-wave  to match the parity of the $\jp$. This is obtained contracting the  $\epsilon^\mu$ polarization vector of $\jp$ with
the momentum of $\bar{\Sigma}^0$.
  Hence, our transition amplitude is
\begin{eqnarray}
t= F \tilde{t}
\end{eqnarray}
with
\begin{eqnarray}
F=\epsilon_\mu p^\mu_{\bar{\Sigma}^0}
\end{eqnarray}
\begin{eqnarray}\label{eq:10}
\tilde{t} &=&\frac{A}{\sqrt{2}}  \left\{G_{K^- p} (M_{\rm inv} (\eta\Lambda)) \, t_{K^- p,\eta\Lambda} (M_{\rm inv} (\eta\Lambda))
- G_{\bar{K}^0 n} (M_{\rm inv} (\eta\Lambda)) \, t_{\bar{K}^0 n,\eta\Lambda} (M_{\rm inv} (\eta\Lambda))  \right\}  \,\no \\
 &+& \frac{B}{\sqrt{2}}  \left\{G_{K^+ \Xi^-} (M_{\rm inv} (\eta\Lambda)) \, t_{K^+ \Xi^-,\eta\Lambda} (M_{\rm inv} (\eta\Lambda))
- G_{K^0 \Xi^0} (M_{\rm inv} (\eta\Lambda)) \, t_{K^0 \Xi^0,\eta\Lambda} (M_{\rm inv} (\eta\Lambda))  \right\}
\end{eqnarray}

When evaluating the sum and average over polarizations  of the $F$ factor we obtain
\begin{eqnarray}
\overline{\sum} \sum |F|^2 =\frac{1}{3} \vec{p}^{~2}_{\bar{\Sigma}^0}
\end{eqnarray}
with $ \bm{p}^2_{\Sigma^0}$ in the $\jp$ rest frame.

 \subsection{The contribution of  $\Lambda(1810) (1/2^+)$ }
\begin{figure}[h!]
\centering
\includegraphics[scale=.85]{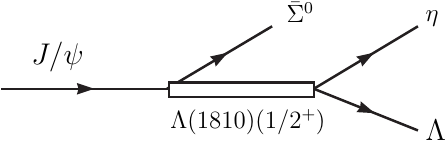}
\caption{Mechanism of $\jp \ra \bar{\Sigma}^0 \eta \Lambda$  through  $\Lambda(1810)(1/2^+)$ excitation.}
\label{fig:2}
\end{figure}

 In Ref. \cite{besexpe}, the fit to the data was done including a  small contribution from the   $\Lambda(1810) \, (1/2^+)$ and we also
include  it here. It gives some small  contribution at higher masses of the $\eta \Lambda$  spectrum and does not distort the shape of the
$\Lambda(1670)$. The $\Lambda(1810)(1/2^+)$ production in $\jp \ra \bar{\Sigma}^0 \Lambda(1810)(1/2^+)$  does not violate parity and
can proceed via  $S$-wave. On the other hand, the $\Lambda(1810)(1/2^+) \ra \eta\Lambda$ requires a $P$-wave  and goes through an operator
$ \vec{\sigma} \vec{p}_\eta$, as the $NN\pi$  vertex. The mechanism for  $\jp \ra \bar{\Sigma}^0 \Lambda(1810)(1/2^+) \ra \bar{\Sigma}^0 \eta \Lambda $ is depicted  in Fig.~\ref{fig:2}.

Since this mechanism does not interfere  with the former one, we sum incoherently the contributions.  The one from Fig.~\ref{fig:2}
can be written  in terms of a $t'$ amplitude such that
\begin{eqnarray}\label{eq:12}
\overline{\sum} \sum  |t'|^2 = C^2 |\frac{1}{M_{\rm inv} (\eta\Lambda)-M_{\Lambda(1810)}+ i  \frac{\Gamma_{\Lambda(1810)} }{2}} |^2 \vec{p}^{~2}_\eta
\end{eqnarray}
with
\begin{eqnarray}
p_\eta =\frac{\lambda^{1/2}(M^2_{\eta\Lambda},m^2_\eta,m^2_\Lambda)}{2 M_{\eta\Lambda}}
\end{eqnarray}

 \subsection{Mass distributions}

 We use the PDG formula \cite{pdg}  updated for the normalization of Mandl and Shaw \cite{mandl}

\begin{eqnarray}\label{eq:14}
\frac{d^2\Gamma}{dM_{12} dM_{23}}=\frac{1}{(2\pi)^3}\frac{2 M_{\bar{\Sigma}^0}\,2 M_\Lambda}{32 M^3_{\jp}}
2 M_{12}\,2 M_{23} \overline{\sum} \sum  (|t|^2 + |t'|^2)
\end{eqnarray}
where we take $\bar{\Sigma}^0 (3)$, $\eta (1)$ and $\Lambda (2)$ labels. In Eq.~\eqref{eq:14} we have
\begin{eqnarray}
\overline{\sum} \sum  |t|^2 = \frac{1}{3} \vec{p}^{~2}_{\bar{\Sigma}^0} |\tilde{t}|^2
\end{eqnarray}
and $|t'|^2$ is given by Eq.~\eqref{eq:12}.

 \subsection{Scattering amplitudes}

 As we can see in Eq.~\eqref{eq:10}, we need the $t_{\bar{K}N,\eta\Lambda}$ and $t_{K\Xi,\eta\Lambda}$ amplitudes. They are calculated
 using the factorized  Bethe Salpeter equation in the coupled channels
 $K^{-} p$, $\bar{K}^{0} n$, $\pi^0 \Lambda$, $\pi^0 \Sigma^0$, $\eta \Lambda$, $\eta \Sigma^0$, $\pi^+ \Sigma^-$, $\pi^- \Sigma^+$,$K^+ \Xi^-$, $K^0 \Xi^0$.
 \begin{eqnarray}\label{eq:16}
 T=[1-V G]^{-1} V
\end{eqnarray}
with $G$ the diagonal meson-baryon loop function, regularised with a cut off and $V_{ij}$ the transition potential between the channels,
evaluated in \cite{angels}. The $f_i$ constants and $q_{\rm max}$ for the channels have been taken from
 \cite{manyu}, where  they are  fitted to the precise data of the
$\Lambda^+_c \to \pi^+ K^- p$ \cite{bellelamb}, and we use them here to be consistent with the data of this reaction  where the
$\Lambda(1670)$  resonance was seen very clearly in the $K^- p$ spectrum.

 \section{Results}
 \begin{figure}[h!]
\centering
\includegraphics[scale=.85]{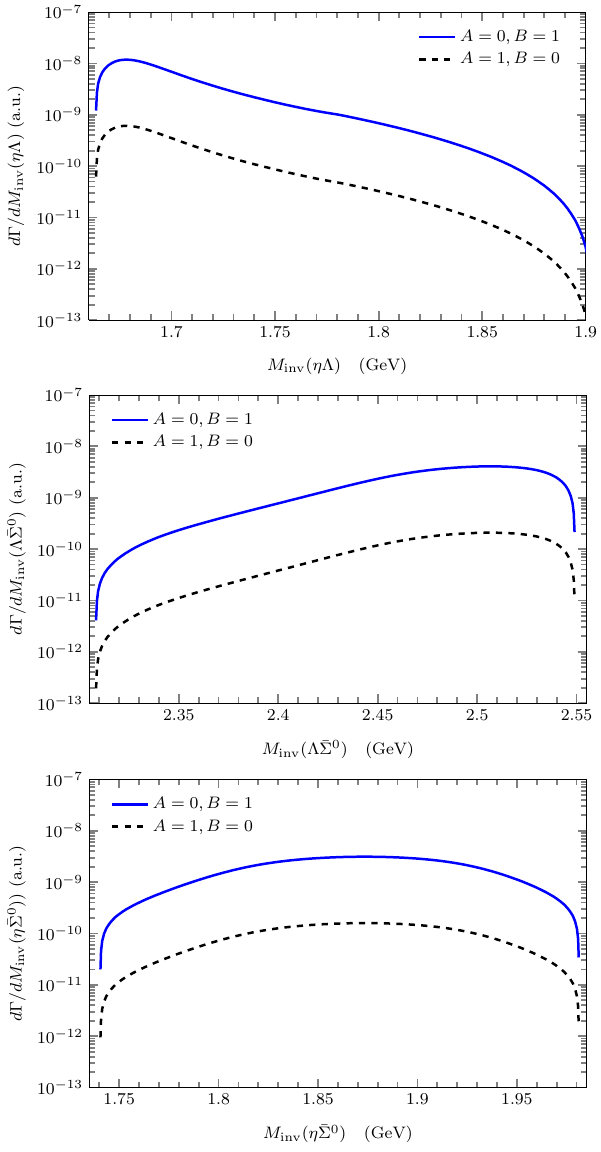}
\caption{The three different mass distributions with $A$ or $B$ equal zero ($C=0$).}
\label{fig:r1}
\end{figure}
 We first look at the relevance of the $A$ and $B$ parameters of  Eq.~\eqref{eq:10} on the mass distributions. This is shown in Fig.~\ref{fig:r1}, where we see that the contribution of the $B$ term, involving the  $K \Xi$ amplitudes has a weight of an order of magnitude  bigger than that of the $A$ term. We learn then that  the  isospin violation is mostly tied to the $K \Xi$ amplitudes, rather than  those of $\bar{K}N$.  The second lesson that we learn  is that  the shapes of the distributions with the $A$ term  alone  or the $B$ term  alone are practically equal, and we find that linear combinations of  the $A$
 and $B$ terms give the same  distributions, up to a global normalization.  This should not be surprising
 since in coupled channels the pole of the resonance comes from ${\rm det} [1-VG]=0$, and the $t_{ij}$ matrices of Eq.~\eqref{eq:16}  go as $A_{ij}/{\rm det} [1-VG]=0$,
with $A_{ij}$ some matrix, hence, all amplitudes share $({\rm det} [1-VG])^{-1}$, and close to the pole the amplitudes are similar up to a  normalization. The same conclusion  is reached if the amplitudes are parameterized
 in terms of the couplings as $T_{ij}\simeq g_i g_j/(M_{\rm inv} -M_R+i \Gamma/2)$.

\begin{figure}[h]
\centering
\includegraphics[scale=.85]{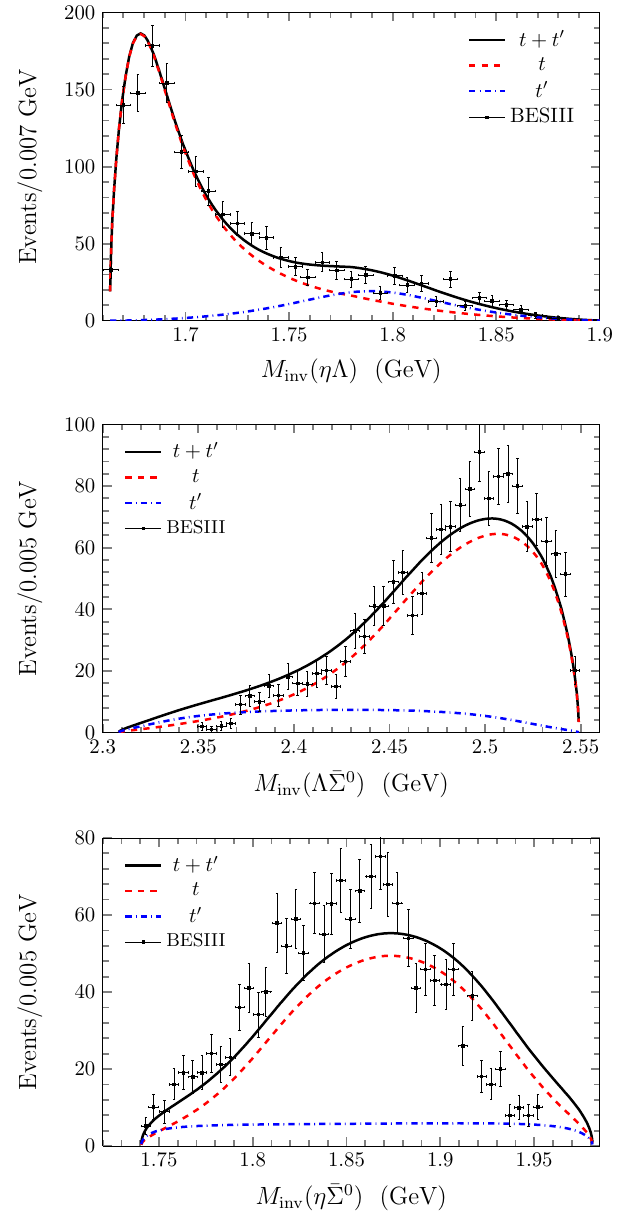}
\caption{results of $\jp\to \Lambda\bar{\Sigma}^0\eta$ decay.}
\label{fig:r2}
\end{figure}

  The next step is to compare  with experiment.
  We carry a best fit to the data taking $A=B$, after knowing that it is the $B$ term that dominates, and the shape is the same with any linear combination of the $A$ and $B$
  terms.\footnote{We take into account in the plots that the $\eta\Lambda$  mass distribution has bins of $0.007\gev$  and the $\Lambda  \bar{\Sigma}^0$  and  $\eta\bar{\Sigma}^0$ bins of $0.005\gev$.}
  Essentially we are determining the relative magnitude of the  $C$ term of Eq.~\eqref{eq:12} and a global normalization.
   We show the results in Fig.~\ref{fig:r2}.  The fit to the data is good, of the same
 quality as the fit performed in the experiment \cite{besexpe}, and the contribution of the $\Lambda(1810)(1/2^+)$ state is also very similar to that in the experimental fit.  The novelty of our work
 consists on showing the dynamical origin of the $\Lambda(1670)$ excitation in this isospin forbidden reaction, the difference of masses of the $\bar{K} N$  and $K \Xi$  charged and neutral modes.

We should mention that the amplitudes that we get with the chiral unitary approach \cite{angels,manyu}
are reliable  in a range of energies that do not exceed $1.8\gev$ in the $\eta\Lambda$  invariant mass. Thus, we do as  it is usually  done \cite{debastiani}, killing adiabatically  the amplitudes after a certain energy.  We take
\begin{eqnarray}
	Gt(M_{\rm cut})=Gt(M_{\rm cut})e^{-\alpha(M_{\rm cut}-M_{\rm cut})} 
\end{eqnarray}
with $M_{\rm cut}=1.8\gev$, $\alpha=5.4\gev^{-1}$ as done in \cite{debastiani}. Note that this simply
reduces the strength of the $\eta\Lambda$ mass distributions  in  a region where its contribution is very small and is dominated by the $\Lambda(1810)(1/2^+)$.

\begin{figure}[h!]
\centering
\includegraphics[scale=.7]{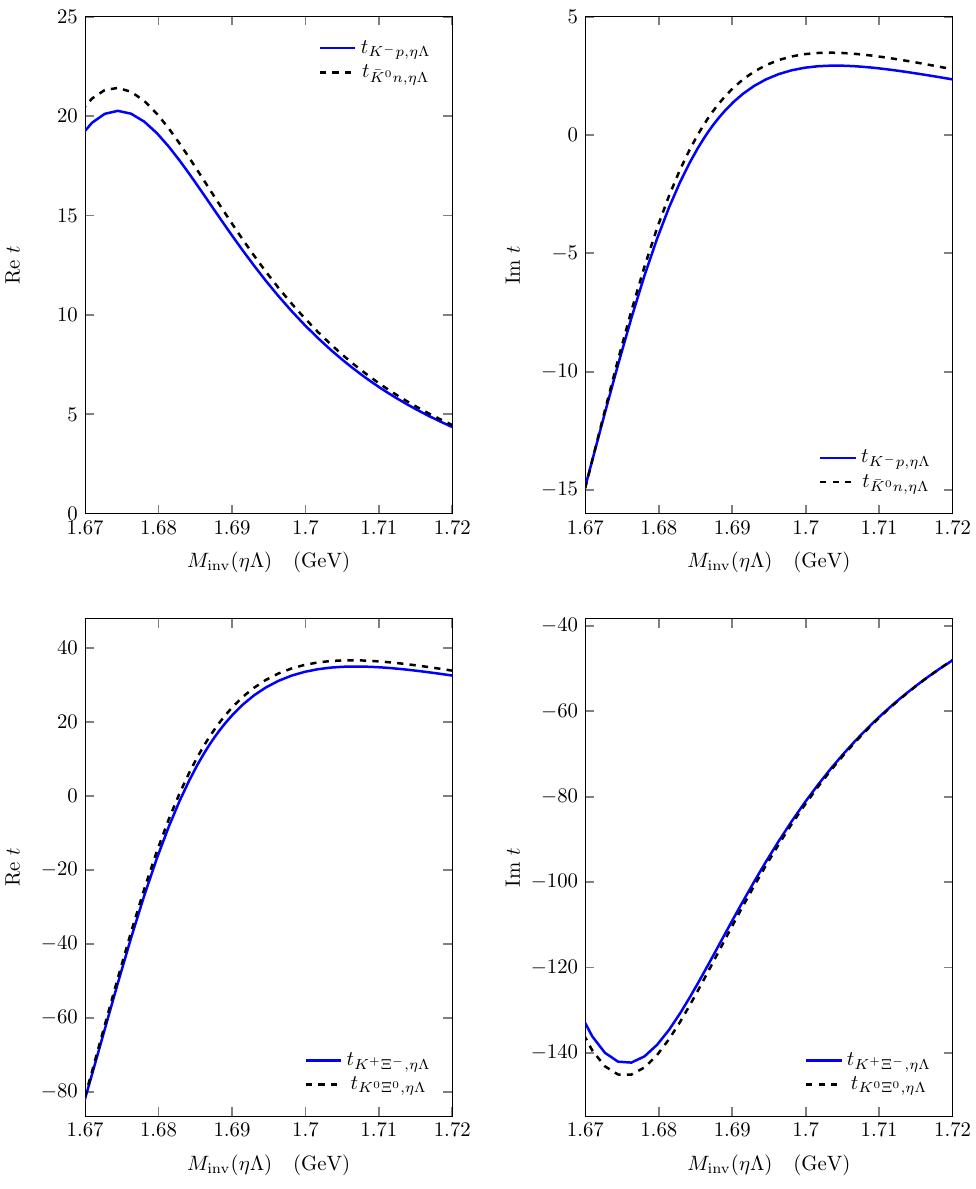}
\caption{The differences between the charged and neutral channels for $t_i$ terms.}
\label{fig:r3}
\end{figure}

\begin{figure}[h!]
\centering
\includegraphics[scale=.7]{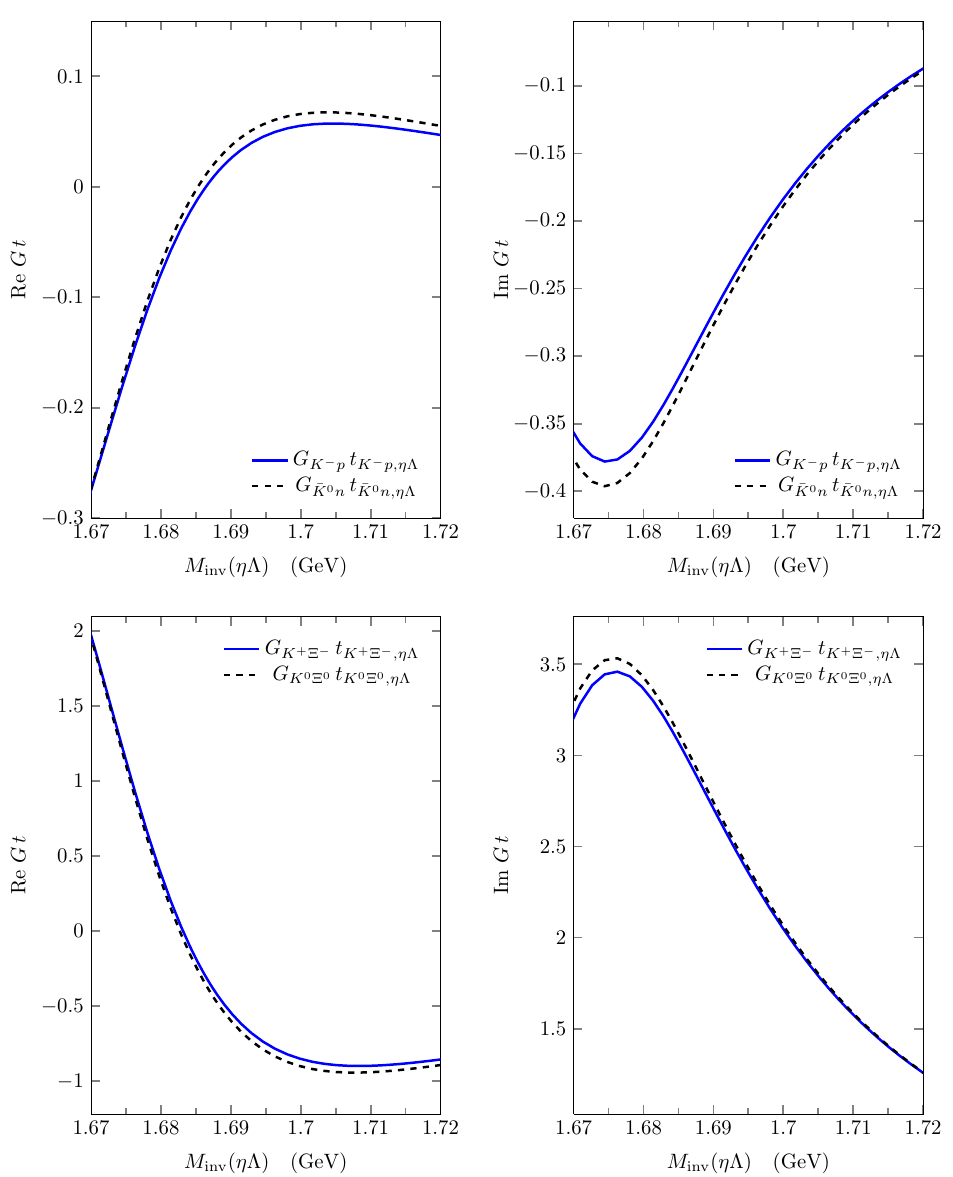}
\caption{The differences between the charged and neutral channels for  $G_i t_i$ terms.}
\label{fig:r4}
\end{figure}

   To get a feeling of the cancellations taking place in the isospin forbidden amplitude, we plot in
Fig.~\ref{fig:r3} the $t_{ij}$ relevant matrices $t_{K^- p,\eta \Lambda}$, $t_{\bar{K}^0 n,\eta \Lambda}$,
$t_{K^+ \Xi^-,\eta \Lambda}$, $t_{{K}^0 \Xi^0,\eta \Lambda}$  appearing in Eq.~\eqref{eq:10}. It is interesting to see that the roles of the
$Re\, t_{\bar{K}N ,\eta \Lambda}$, $Im\, t_{\bar{K}N ,\eta \Lambda}$ and the
$Re\, t_{K \Xi,\eta \Lambda}$, $Im\, t_{K \Xi,\eta \Lambda}$  are exchanged, but this happens often  in coupled channels in inelastic amplitudes.
 The interesting thing to observe  is that the magnitudes $Gt$
are remarkable similar  up to a change of sign and a normalization, such that combinations with $A$ and $B$ with opposite sign
in  Eq.~\eqref{eq:10} lead to larger decay  widths than if $A$ and $B$ have same sign. In any case, the figures tell us why the terms in Eq.~\eqref{eq:10}  with the  $t_{K \Xi,\eta \Lambda}$  amplitudes dominate. Indeed $|Im\, t_{K \Xi ,\eta \Lambda}|$  is about an order of magnitude bigger than
$|Im\, t_{K \Xi,\eta \Lambda}|$.
 We also observe in Fig.~\ref{fig:r3}  that  there are small differences between the charged and neutral
 channels in the $t$ matrix.
 It is also interesting to plot the results for $Gt$ entering the evaluation of Eq.~\eqref{eq:10}.
 We show the results in  Fig.~\ref{fig:r4}.  It is interesting to see that Re $G t$ and Im $G t$ are now
 similar in shape, up to a global sign. This is a consequence  of the fact that Im $G_{\bar{K}N}$ dominates $G_{\bar{K}N}$  and hence
 in $G_{\bar{K}N} t_{\bar{K}N,\eta \Lambda}$  the roles of the real and imaginary part are now reversed with respect to $t_{\bar{K}N,\eta \Lambda}$.
  The differences observed in $Gt$ for the different channels
 are responsible for the isospin violation in the reaction studied.

\section{Conclusions}
   We have studied theoretically the $J/\psi \to \Lambda\bar \Sigma^0 \eta $ reaction, which has been recently measured by the BESIII collaboration, in which the $\Lambda(1670)$ resonance is excited and practically accounts for all the strength of the mass distributions. Taking advantage that the
$\Lambda(1670)$ resonance  is one of those generated by coupled channels interaction using the chiral unitary approach, we have studied the reaction from this perspective and looked at the dynamical source of the isospin violation. Like in many other reactions, this violation is blamed on the effect of loops involving kaons, mostly in the $K \Xi$  loops, due to non cancellation of terms in isospin allowed amplitudes because of the different masses of the charged and neutral particles belonging to the same isospin multiplets. In order to evaluate the amplitudes, we look first at the $J/\psi$ decay to  $\bar \Sigma^0$ and a pair of pseudoscalar-baryon states, both belonging to SU(3) octets. For this we consider the $J/\psi$ as an SU(3) singlet and look for the dominant structures of an antibaryon, a baryon and a pseudoscalar, coupling to an SU(3) singlet. Taking advantage of the couplings obtained in the study of the meson baryon chiral unitary approach, the isospin violation is linked at the end to the effect of the $K^- p, \bar K^0 n$ and $K^+ \Xi^-,K^0 \Xi^0$ channels, and the study done shows that the $K^+ \Xi^-,K^0 \Xi^0$ channels are mostly responsible  for the strength of the mass distributions.

   Up to a global normalization, that we cannot determine with just the SU(3) symmetry argument used for the $J/\psi$ decay to  $\bar \Sigma^0$ and a pair of pseudoscalar-baryon states, the mass distributions that we obtain are in good agreement with the experimental mass distributions, in particular when adding the contribution of the $\Lambda(1810) (1/2^+)$ that we have just fitted to the experiment and is very similar to the one obtained in the experimental fit. In summary, we have obtained an appealing and very plausible explanation for this interesting isospin violating reaction, based on previous findings of meson baryon interaction provided by the chiral unitary approach.

\begin{acknowledgments}

L. R. Dai would like to acknowledge some useful discussions with Xiao-shen Kang.
This work is partly  supported by the National Natural Science
Foundation of China under Grant No. 12575082.
This work was also partly supported by the National Key R\&D Program of China (Grant No. 2024YFE0105200), the Natural Science Foundation of Henan (Grant No. 252300423951), the National Natural Science Foundation of China (Grant No. 12475086), and the Zhengzhou University Young Student Basic Research Projects for PhD students (Grant No. ZDBJ202522). Wen-Tao Lyu acknowledges the support of the China Scholarship Council.
This work is also partly supported by the Spanish Ministerio de Economia y Competitividad (MINECO) and European FEDER funds under Contracts No. FIS2017-84038-C2-1-PB, PID2020-112777GB-I00, and by Generalitat Valenciana under contract PROMETEO/2020/023. This project has received funding from the European Union Horizon 2020 research and innovation program under the program H2020-INFRAIA-2018-1, grant agreement No. 824093 of the STRONG-2020 project.

\end{acknowledgments}

\end{document}